\documentclass[3p,number]{elsarticle}
\usepackage{amsmath,amssymb,mathrsfs}
\usepackage{latexsym}
\usepackage{CJK}
\usepackage{graphicx}
\usepackage{indentfirst}
\usepackage{listings}
\usepackage{xcolor}
\usepackage{booktabs}
\usepackage{stmaryrd}
\usepackage{multirow}
\usepackage{verbatim}
\usepackage{makecell}
\usepackage{lineno,hyperref}
\usepackage{longtable}
\usepackage{hyperref}
\usepackage{upgreek}
\usepackage{mathtools}
\usepackage{enumitem}
\usepackage{float}
\usepackage{graphicx}
\usepackage{subfig}
\usepackage{bm}   
\newcommand{\matr}[1]{\bm{#1}}

\hypersetup{colorlinks,breaklinks,           linkcolor=blue,urlcolor=blue,anchorcolor=blue,citecolor=blue}

\begin{document}

\title{Simulations of grain growth in tungsten armor materials under ARC plasma edge operation conditions using an integrated plasma-edge/materials model}

\author[ucla]{Jinxin Yu\corref{cor1}}
\author[lanl]{Nithin Mathew}
\author[utk]{Sophie Blondel}
\author[utk]{Ane Lasa}
\author[cfs]{Jon Hillesheim}
\author[cfs]{Lauren Garrison}
\author[utk]{Brian D.\ Wirth}
\author[ucla]{Jaime Marian\corref{cor2}}

\address[ucla]{$^1$Department of Materials Science and Engineering, University of California Los Angeles, Los Angeles, CA 90095, USA}
\address[lanl]{$^2$Group XCP-5, X Computational Physics Division, Los Alamos National Laboratory, Los Alamos, NM 87544, USA}
\address[utk]{$^3$Department of Nuclear Engineering, University of Tennessee-Knoxville, Knoxville, TN 37996, USA}
\address[cfs]{$^4$Commonwealth Fusion Systems, USA}
\cortext[cor1]{jinxinyu@g.ucla.edu}
\cortext[cor2]{jmarian@ucla.edu}

\vspace{10pt}

\begin{abstract}
An integrated model of grain growth deuterium-exposed tungsten polycrystals, consisting of a two-dimensional vertex dynamics model fitted to atomistic data, has been developed to assess the grain growth kinetics of deuterium-exposed polycrystalline tungsten (W). The model tracks the motion of grain boundaries under the effect of driving forces stemming from grain boundary curvature and differential deuterium concentration accumulation. We apply the model to experimentally-synthesized W polycrystals under deuterium saturated conditions consistent with those of the ARC concept design, and find fast grain growth kinetics in the material region adjacent to the plasma (at 1400 K, $<$100 seconds for full transformation), while the microstructure is stable deep inside the material (several days to complete at a temperature of 1000 K). Our simulations suggest that monolithic W fabricated using conventional techniques will be highly susceptible to grain growth in the presence of any driving force at temperatures above 1000 K.
\end{abstract}

%

 
%
\maketitle
\vspace{2pc}



\section{Introduction}\label{sec:intro}

Structural materials for fusion energy systems must be designed to face extreme conditions during reactor operation \cite{bolt2002plasma,raffray2010high,antusch2017refractory,hancock2018refractory}.
In particular, divertors in magnetic fusion devices will be subjected to nominal heat fluxes exceeding 10 MW$\cdot$m$^{-2}$ \cite{bolt2004materials,stork2014developing,linsmeier2017development,kuang2020divertor} (and even higher during sporadic discharge events known as \emph{edge localized modes}, which can reach heat loads of up to 100 MW$\cdot$m$^{-2}$ \cite{guillemaut2015ion,loewenhoff2011evolution,leonard2014edge}). Tungsten (W) has emerged as the main candidate plasma-facing component (PFC) material due to a set of favorable properties, including a high melting point (3695 K), high thermal conductivity (173 W$\cdot$m$^{-1}$$\cdot$K$^{-1}$ at room temperature), low tritium storage, low coefficient of thermal expansion ($4.3\times10^{-6}$ K$^{-1}$), low neutron activation, low erosion rate and high-temperature yield strength. However, the use of W is limited by a narrow operational temperature window, with its upper limit set by recrystallization, and its lower one by a generally unacceptably-high brittle-to-ductile (DBTT) transition temperature \cite{naujoks1996tungsten,abernethy2017predicting} (which can range between room temperature for high-purity, cold-rolled specimens \cite{reiser2016ductilisation,bonnekoh2019brittle,brunner2000analysis} to over 500$^\circ$C for conventionally forged specimens \cite{Kim2014,butler2018mechanisms,ren2019investigation}). 
In addition, due to the intermittent --but cyclic-- nature of plasma kinetic behavior during fusion reactor operation, it is also important to consider elements of thermal fatigue in the mechanical performance of W components \cite{mcelfresh2024fracture,pintsuk2013qualification,pintsuk2015characterization,loewenhoff2012tungsten,wang2016thermal,fukuda2020effect,shah2021numerical,ma2023high}. 

The DBTT of conventional polycrystalline tungsten depends on the material microstructure, and can be lowered by plastic deformation to values even below room temperature \cite{reiser2012tungsten,reiser2016ductilisation,pantleon2021thermal}. Pre-straining by, e.g., cold working or rolling is the basis for the processing of the so-called `fusion-grade' tungsten \cite{terentyev2017mechanical,yin2018tensile}. However, the deformation structures resulting from low-temperature deformation are thermally unstable, and predispose the material to a set of restoration processes during operation at high temperatures: recovery, recrystallization, and grain coarsening. The unavoidable occurrence of one or more of these processes during fusion device operation ultimately brings W back to its inherently brittle undeformed state, effectively setting practical limits on the material's lifetime. Recrystallization and grain growth are enabled by grain boundary (GB) motion, driven by differential driving forces resulting from chemical potential jumps across the boundary \cite{mcelfresh2023initial}, or from the anisotropy of the GB energy itself \cite{rohrer2023grain}. A natural strategy to delay grain coarsening processes is then to enable Zener pinning by selective alloying and GB segregation \cite{lang2021recrystallization,gietl2022neutron} (notwithstanding other detrimental processes that might result from alloying). For example, the recrystallization temperature limit in W-3Re alloys doped with K was seen to increase beyond 850$^\circ$C relative to that of pure W \cite{gietl2022neutron}.

Under operation, the divertor will be inevitably exposed to hydrogen isotopes (H/D/T) from the plasma, resulting in the penetration of H atoms inside the material and interactions with its internal microstructure. In particular, grain boundaries are known to trap H ions with energies ranging between 0.8 and 1.2 eV \cite{piaggi2015hydrogen,diaz2022direct,zheng2025role}, a process which has been seen to decrease their intrinsic mobility \cite{mathew2022interstitial,yu2015effect} and alter the equilibrium concentration profiles that would be expected from Fickian kinetics. However, the collective effect of trapped and free H isotopes on microstructural evolution during grain growth is not well understood. Past studies \cite{shah2021recrystallization,buzi2017response,saad2018tungsten,mannheim2018modelling} have shown no clear difference between pristine and H-loaded W specimens at temperatures near the recrystallization limit. 
Still, characterization of internal H concentration profiles during H-plasma expoure is exceedingly challenging, and the quantitative correlation between H-isotope concentration and grain growth remains to be established.
It is thus crucial to gain a detailed understanding of the processes leading to microstructural changes of W-based components exposed to hydrogen isotopes. 

In this paper, we carry out vertex dynamics simulations of grain growth in plasma-exposed polycrystalline W by adapting an existing model for recrystallization \cite{mcelfresh2023initial} to the particular conditions expected in \emph{Commonwealth Fusion Systems}' next-of-a-kind fusion concept ARC \cite{segantin2020exploration}. The ARC is in an early design stage, so the choice of the surface and substrate material, as well as the details of the geometric design and allowable plasma parameters are still under development. These simulations are thus intended as a guide to setting reasonable limits and expectations on materials performance under ARC's planned operational conditions.
As boundary conditions, we use depth-resolved temperature distributions and D-T concentrations recently calculated by an integrated plasma-materials interaction model. The model captures boundary plasma conditions as obtained by the \texttt{UEDGE} code\cite{wigram2019performance}, ion-surface interactions by \texttt{F-TRIDYN}, and sub-surface gas dynamics by \texttt{Xolotl} \cite{maroudas2016helium}. This integrated approach was then used to simulate the exposure in ten locations along the outer divertor target in detached plasma conditions \cite{lasa2024development}. 

The paper is organized as follows. First, a review of the theory behind grain boundary evolution models and the implementation adopted in this work are provided in Sec.\ \ref{meth}.
The results of the paper are presented in Sec.\ \ref{res}, including the selection of the microstructure from characterization maps and grain simulations under several different scenarios.
We provide a discussion of our findings in Sec.\ \ref{disc}, and finish with the main conclusions of the paper in Sec.\ \ref{conc}.

\section{Theory and methods}\label{meth}

\subsection{Vertex model for grain boundary kinetics}

Our model of grain boundary evolution is based on a two-dimensional (2D) vertex model of the polycrystal microstructure \cite{mcelfresh2023initial}. The model consists of \emph{physical} nodes, representing triple junctions (TJ), and \emph{virtual} nodes connected as a piecewise discretization of a grain boundary. A summary of the features of the model is provided in the next section.

\subsubsection{Virtual node evolution equations:}

The motion of a grain boundary (GB) is defined by a generalized displacement $\mathbf{u} = (u_1 , u_2 , u_3 )$ with component $u_1$ representing motion in the direction of the GB normal, and $u_2$ and $u_3$ representing in-plane motion. The GB velocity, $\vec{v}=\dot{\vec{u}}$, responds to a set of driving forces defined by the vector $\vec{f}$:
\begin{equation}
\vec{v}=\matr{M}\vec{f}=\begin{pmatrix}
M_{11} & M_{12} & M_{13} \\
M_{12} & M_{22} & M_{23} \\
M_{13} & M_{23} & M_{33} 
\end{pmatrix}
\begin{pmatrix}
\psi \\
 \sigma_{12} \\
  \sigma_{13}
\end{pmatrix}\label{pepei}
\end{equation}
where $\matr{M}$ is the mobility matrix. The first component of $\vec{f}$, $f_1=\psi$, includes all `chemical' sources of motion. The second and third components of $\vec{f}$ are $f_2 = \sigma_{12}$ and  $f_2 = \sigma_{13}$, which are shear stresses along the $x_2$ and $x_3$ directions in the GB plane. These stresses lead to so-called \emph{coupled shear boundary} motion through the corresponding mobility components $M_{12}$, $M_{23}$, and $M_{13}$. 

In 2D, eq.\ \eqref{pepei} is reduced to:
\begin{eqnarray*}
v_1=M_{11}\psi+M_{12}\sigma_{12}\\
v_2=M_{21}\psi+M_{22}\sigma_{12}
\end{eqnarray*}
For zero stress conditions (see Secs.\ \ref{beta} and \ref{tj2}) and assuming a decomposition of the GB normal $\vec{n}$ as $n_1=\cos\phi$ and $n_2=\sin\phi$ (where $\phi$ is the polar angle formed by the GB normal relative to a 2D cartesian frame of reference), the corresponding mobilities can be expressed as:
\begin{eqnarray*}
M_{11}=M(\theta,T)n_1\\
M_{21}=M(\theta,T)n_2
\end{eqnarray*}
where $M_{11}$ captures motion of a GB vertex along the GB normal, and $M_{21}$ represents its motion in the GB plane. $M(\theta,T)$ is defined in Sec.\ \ref{sec:mob}. Motion governed by $M_{11}$ responds to a mechanical driving force, while motion defined by $M_{21}$ reacts to `configurational' driving forces. 
In this work, we disregard the latter, which leads to the following simplified equation of motion:
\begin{equation}\label{GB_velocity_2D}
    v_1 = M_{11}\psi
\end{equation}

\subsubsection{Chemical driving force for GB motion:}

\begin{figure}[htb]
\centering
\includegraphics[height=0.35\linewidth]{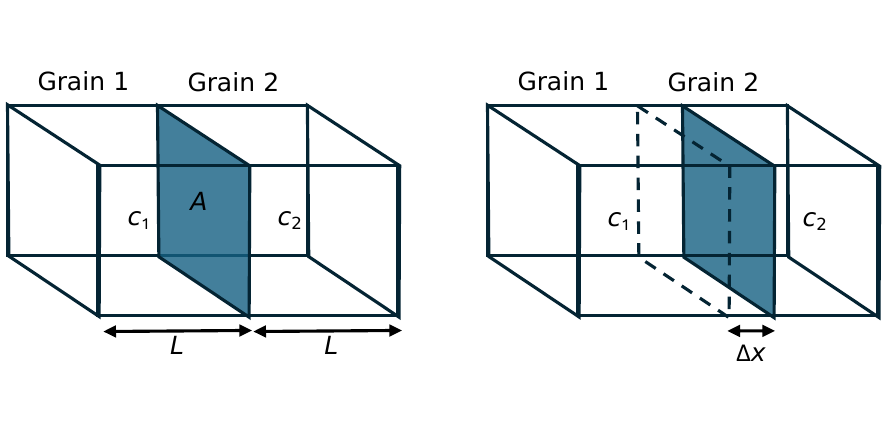}
\caption{\label{fig:D_GB_motion}
Bi-crystal structure used to derive the driving force on the GB due to a deuterium concentration jump.}
\end{figure}

The components of $f_{1}=\psi$ in eq.\ \eqref{pepei} that contribute to GB motion include:
\begin{itemize}
    \item[(i)] \underline{Curvature-driven motion:}
\begin{equation}
\vec{f}_\gamma = -\frac{\gamma(\theta)}{R}\vec{n}
\label{driving}
\end{equation}
where $R$ is the radius of curvature of the approximate circle representing the GB at a given instant in 2D, and $\gamma(\theta)$ is the misorientation-dependent GB energy ($\theta$ is the misorientation angle), $\vec{n}$ denotes the unit normal vector of the GB surface, oriented outward in the direction of the local curvature. 
 \item[(ii)] \underline{Differential deuterium concentration}: If we consider the ideal bi-crystal shown in Fig.~\ref{fig:D_GB_motion}, composed of two identical grains of length $L$, a GB area $A$, and each containing a deuterium concentration $c_1$ and $c_2$, a force to move the GB towards the grain with the higher deuterium concentration will exist. This force can be derived by calculating the total energy of the deuterium in solution in the bi-crystal:
\begin{equation}
    E^{\mathrm{tot}}_{i} = AL \left(c_1 +c_2\right)\Delta h_{\text{D}}
    \label{eq:diff-deut}
\end{equation}
where $\Delta h_{\text{D}}$ is the enthalpy of solution (formation enthalpy) of deuterium in tungsten per atom, taken here as constant.
After the GB moves by a distance $\Delta x$ into grain 1, the current total energy of bi-crystal is
\begin{align}\label{current_energy}
    E^{\mathrm{tot}}_{f}=A \left(c_1 (L + \Delta x) + c_2 (L - \Delta x)\right)\Delta h_{\text{D}} 
\end{align}
For infinitesimal displacements of the GB, $c_1$ and $c_2$ can be assumed to remain unchanged after the GB moves, the driving force for the GB motion resulting from deuterium concentration jump is equal to:
\begin{equation}\label{force_D_point}
    f_{\text{D}}^{\text{tot}} = -\frac{\Delta E^{\text{tot}}}{\Delta x}=-\frac{E_{\text{f}}^{\text{tot}}-E_{\text{i}}^{\text{tot}}}{\Delta x}=A\left(c_1 - c_2\right)\Delta h_{\text{D}} 
\end{equation}
The above expression is for a true force acting on a point. 
Dividing this force by the GB area $A$ gives the capillary pressure due to the concentration
differential across the interface:
\begin{equation}
    \vec{f}_{\text{D}}=\Delta h_{\text{D}}(c_1 - c_2) \vec{n}=\Delta h_{\text{D}}\Delta c\vec{n}
\end{equation}
where $\Delta c$ is the deuterium concentration jump across the GB and $\vec{n}$ is the unit normal vector of the GB. 
A positive force is taken for $c_1 > c_2$, i.e., with the GB moving in the direction of grain with the higher deuterium concentration.

\end{itemize}

Combining both forces above, the total chemical driving force $\vec{f}_\psi$ experienced by a GB is: 
\begin{equation}\label{chemical_GB_force}
    \vec{f}_{\psi} = \vec{f}_\gamma + \vec{f}_{\text{D}}=\left[-\frac{\gamma(\theta)}{R}+\Delta h_{\text{D}}\Delta c\right]\vec{n}.
\end{equation}
which is used in eq.\ \eqref{GB_velocity_2D} in its scalar form $\psi=\|\vec{f}_{\psi}\|$. The function $\gamma(\theta)$ is defined in Sec.\ \ref{sec:gb-en}. For $\Delta h_{\text{D}}$, we take a value of 1.4 eV consistent with DFT calculations \cite{hodille2017simulations,yu2020understanding}.

\subsubsection{Shear-coupled motion:}\label{beta}

The shear-coupling factor (the ratio of the shear velocity to the GB migration velocity) has two components $\beta_k = v_k /v_1$ corresponding to orthogonal shear directions contained in the GB facet. This factor can be obtained by measuring $v_1$ and $v_k$ under an applied shear stress $\sigma_{1k}$, i.e., $\beta_k =M_{kk}/M_{1k}$, which can be done directly using MD simulations \cite{PhysRevLett.109.095501,PhysRevB.78.064106,chen2019grain,thomas2017reconciling}.

In the present study no stress is applied and thus there is no direct shear coupled GB motion from external stresses. Rather, it is intrinsic (i.e., `regular') GB motion that induces internal shear stresses. As discussed in Sec.\ \ref{tj2}, these stresses transfer to triple junctions, creating the possibility for twin or dislocation nucleation into the grain \cite{thomas2019disconnection}. However, at present we assume that triple junctions have sufficiently high mobilities to relieve these stresses as they appear.


\subsubsection{Triple junction motion:}\label{tj2}

The role of triple junctions (TJ) in controlling grain growth in annealed polycrystalline materials was recognized in early theoretical and simulation studies \cite{gottstein2005triple,gottstein2002triple,gottstein2004validity}. Specifically, when grain growth in a polycrystal is governed by the mobility of triple junctions, the kinetics deviates from the classical Johnson-Mehl-Avrami \cite{price1990use} --or its variant, the Mullins-Neumman \cite{gottstein2004validity}-- evolution law.

Triple junctions move according to the following viscous law:
\begin{equation}
\vec{v}_{\rm TJ}={\cal M}_{\rm TJ}\sum_{k=1}^3\gamma_k\vec{t}_k
\label{tj1}
\end{equation}
where the index $k$ refers to the three GB that converge on the TJ and ${\cal M}_{\rm TJ}$ is the triple junction mobility. $\gamma_k$ is the energy per unit area of GB $k$. The three forces from the GBs converging into the TJ are oriented along the GB tangents $\vec{t}_k$.  
In equilibrium, $\vec{v}_{\rm TJ}$ is zero and if the GB energy is independent of misorientation, one gets the dihedral angle condition (Herring relation). Indeed, eq.\ \eqref{tj1} implies that grains with dihedral angles of less than $90^\circ$ are always unstable and will eventually disappear.

When the ratio $\Lambda=L{\cal M}_{\rm TJ}/M_{\rm GB}$ is large ($L$ is the grain size: note that the units of the TJ and GB mobilities differ by one length dimension), the dihedral angle condition is maintained because the TJ can adjust instantaneously to the motion of the GBs. When $\Lambda$ is small, steady state dihedral angles that deviate from the static solution may be seen, but those are physical and may ultimately control microstructural evolution. In other words, if triple junctions can move sufficiently rapidly to remain in equilibrium with respect to the GBs, then the microstructure evolves by pure chemical forces with fixed angle junctions. This type of motion may be viewed as a two-step process whereby (i) GBs migrate in accordance with eq.\ \eqref{pepei}, thereby pulling the TJs out of equilibrium, and (ii) the TJs migrate to restore the equilibrium angles following eq.\ \eqref{tj1}. Experimental observations and simulation show that TJs do indeed have finite mobility, as indicated by steady-state dihedral angles of migrating TJs that differ from their equilibrium values \cite{gottstein2001grain,gottstein2002triple,gottstein2005triple,thomas2019disconnection}.

Quantitative assumptions to define the value of ${\cal M}_{\rm TJ}$ and $\Lambda$ are provided in Sec.\ \ref{sec:tjmob}.

\subsection{Deuterium concentration jump across a grain boundary}\label{sec:hjump}

\begin{figure}[htb]
    \centering   \includegraphics[width=0.3\linewidth]{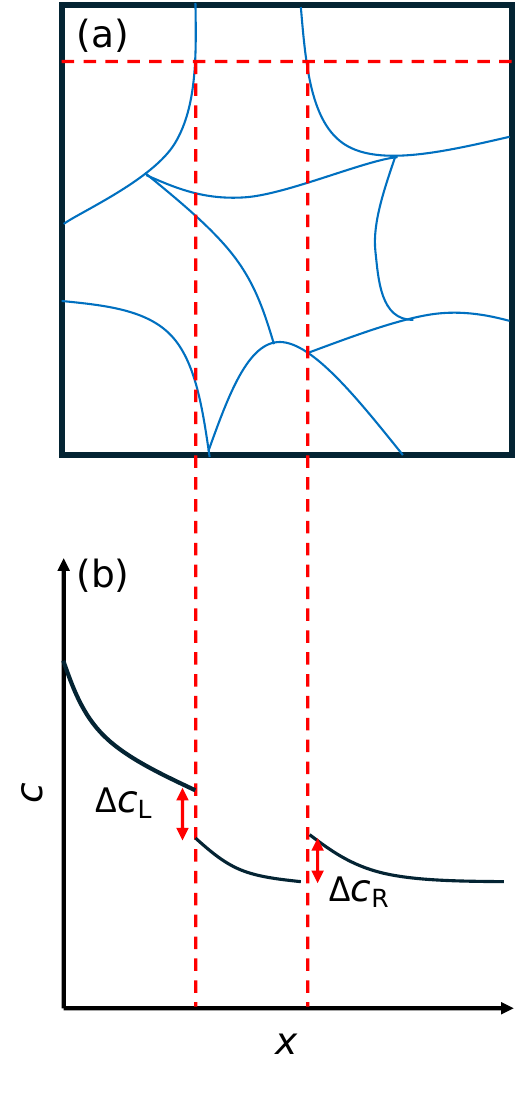}
    \caption{\label{fig:low_angle_GB}
    (a) The schematic of two-dimensional polycrystal model;
    (b) The schematic of concentration discontinuity across the GB.}
\end{figure}

The presence of grain boundaries is known to alter impurity concentration profiles relative to those in a homogeneous (e.g., single crystal) environment. Studies have revealed discontinuous segregation patterns across twin boundaries \cite{Dingreville2023}, thin films \cite{Lu2019}, and grain boundary complexions \cite{garg2021segregation}. When the solute concentration is uniform, segregation enrichment or depletion are common in polycrystals \cite{Karlsson1988,Guin2023}. However, when the concentration field is biased (e.g., subjected to a gradient), `jumps' in solute concentration can be seen across boundaries without changing the concentration gradient. These jumps lead to differential solute concentrations that result in a driving force for GB motion. The process is schematically shown in Figure \ref{fig:low_angle_GB}, where a solute concentration profile is superimposed on a generic polycrystalline system, leading to jumps in concentration across each GB. 
In this section we derive the relationship between the solute concentration differential, $\Delta c$, and the GB misorientation, $\theta$.

Our approach is based on the consideration of an excess `free' volume, $\delta V$, of a GB relative to the bulk, an idea going back to solute segregation models by Aaron and Bolling in 1972~\cite{aaron1972free}. In the absence of more accurate ways to obtain it, Wolf proposed a direct correlation between $\delta V$ and the GB misorientation following the Shockley-Read expression for GB energies \cite{wolf1989correlation}. However, the excess volume can now be readily obtained from atomistic simulations as:
\begin{equation}\label{excess_volume}
    \delta V = \frac{V_{\text{GB}}-N\Omega_a}{2A}
\end{equation}
where $V_{\text{GB}}$ is the volume of a simulation box containing a periodic bi-crystal (i.e., two equivalent grain boundaries with equal surface area $A$), $N$ is the total number of atoms in the simulation, and $\Omega_a$ is the atomic volume (equal to $a_0^3/2$ in BCC metals, where $a_0$ is the lattice constant). Equation \eqref{excess_volume} compares the actual volume of a system containing a fixed number of atoms and two grain boundaries with the volume that those atoms would occupy in a bulk system. To remove the scale effect associated with the arbitrary value of $N$, $\delta V$ is normalized by the GB area (and thus expressed in units of length). Note that the excess volume can be positive or negative, depending on where the tilt angle is measured from. 

The concentration jump across a GB with an associated excess volume of $\delta V$ can be obtained by assuming that the impurity concentration in the bulk experiences a In the case of deuterium atoms, the same number of particles, $n_{\rm D}$, is now distributed over a different volume than in the bulk. The concentration prior to encountering a GB, $c_i$, can be written as:
$$c_i=\frac{n_{\rm D}}{V_i}$$
where $V_i=wA$ is the volume of a region adjacent to the GB with a width equal to $w$. At the GB, the total available volume changes to:
$$V_f=V_i\pm A\delta V$$
where $A=L^2$ is the GB surface area. The deuterium concentration after the GB is now:
$$c_f=\frac{n_{\rm D}}{V_f}=\frac{n_{\rm D}}{V_i+A\delta V}=\frac{n_{\rm D}}{A\left(w\pm\delta V\right)}$$
The concentration jump, $\Delta c=c_f-c_i$, is then:
\begin{align}
    \Delta c=c_f-c_i&=\frac{n_{\rm D}}{A\left(w\pm\delta V\right)}-\frac{n_{\rm D}}{wA}=c_i\left(\frac{1}{1\pm\delta V/w}-1\right)\nonumber\\
    &=c_i\frac{\delta V}{w\pm\delta V}\approx\pm c_i\frac{\delta V}{w}\label{eq:width}
\end{align}
where we have assumed that $\delta V\ll w$. Thus, the excess volume leads to a correction in the concentration profile that scales with the concentration measured at the GB on the side closest to the exposed wall of the material.



\subsection{Physical parameters}\label{sec:phys}

\subsubsection{Grain boundary energies:}\label{sec:gb-en}
\begin{figure}[htbp]
    \centering
    \subfloat[Grain boundary energy\label{fig:GBEnergy}]{
        \includegraphics[width=0.45\textwidth]{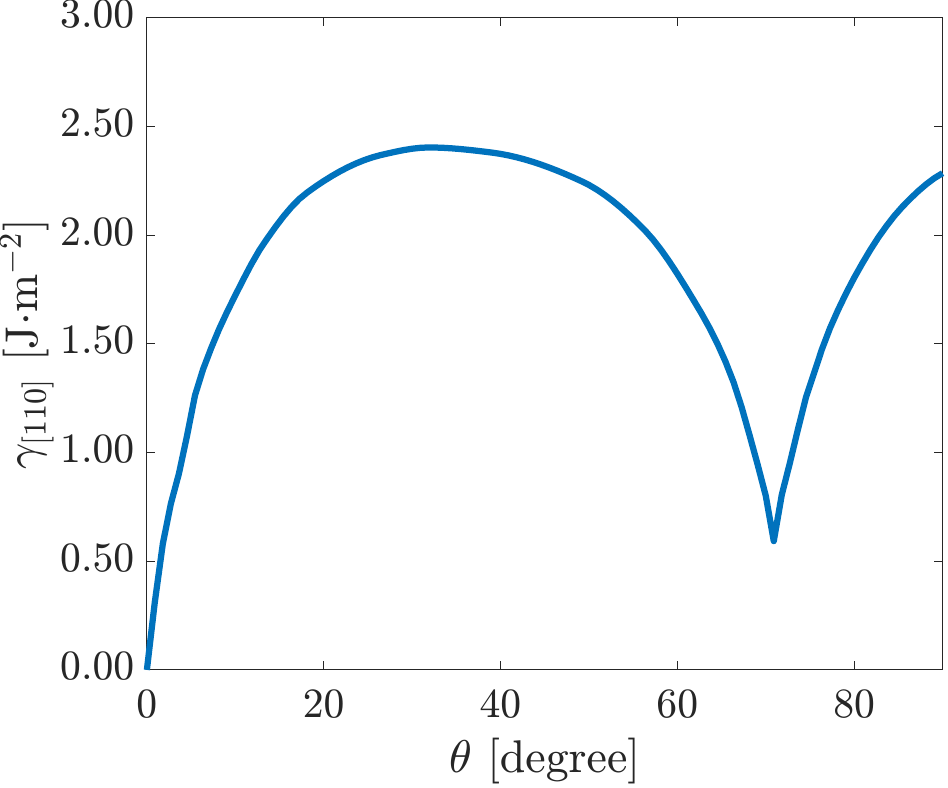}
    }
    \hspace{0.02\textwidth} 
    \subfloat[Excess volume\label{fig:GBexcess}]{
        \includegraphics[width=0.47\textwidth]{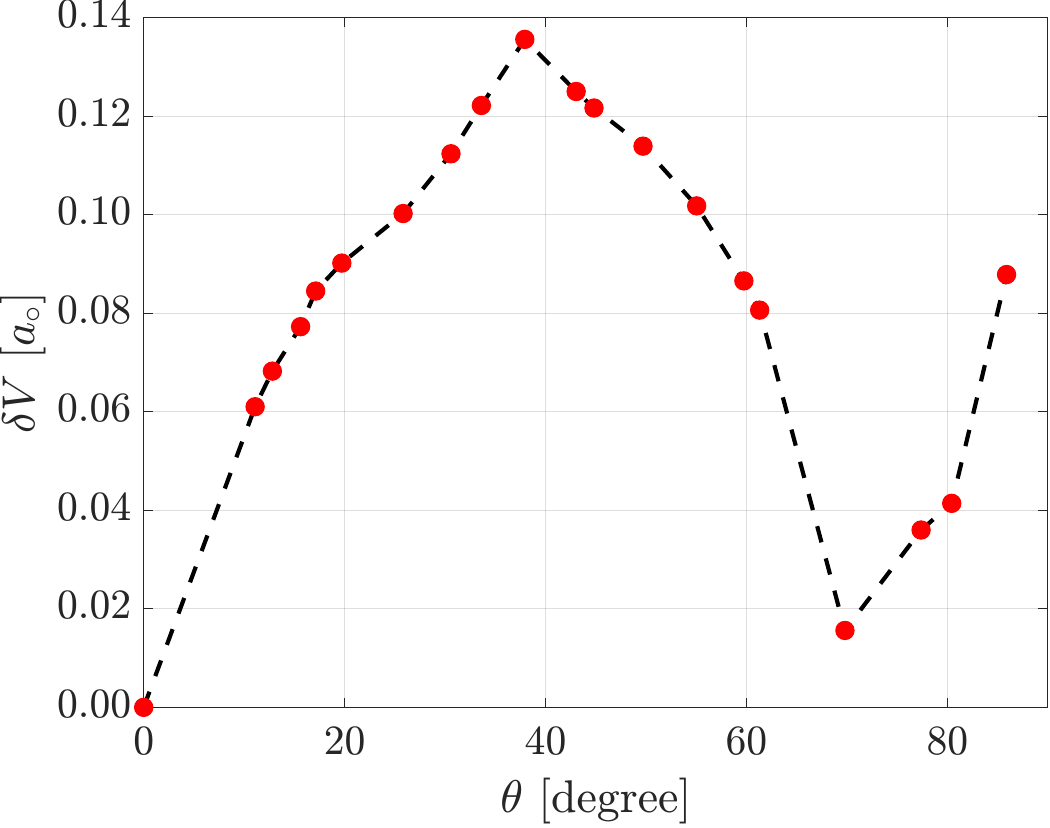}
    }

    \caption{(a) $[110]$ symmetric tilt grain boundary energies versus misorientation angle $\theta$ (from~\cite{chirayutthanasak2022anisotropic}); (b) Excess volume of $[110]$ symmetric tilt GB versus misorientation angle (from ref.~\cite{zheng2023structure}).}
    \label{fig:GBEnergyExcess}
\end{figure}

In this paper, we take GB energies with a non-monotonic dependence on misorientation obtained using atomistic calculations. Specifically, we consider energies for $[110]$ symmetric tilt boundaries calculated using semi-empirical potentials \cite{frolov2018structures,feng2015energy,chirayutthanasak2022anisotropic} validated with electronic structure calculations \cite{setyawan2014ab,setyawan2012effects}. The $\gamma_{[110]}$-$\theta$ function for these family of GB is presented in Figure~\ref{fig:GBEnergy}. 


The data displayed in the figure are then used to fit a set of Read-Shockley-Wolf functions (RSW) functions \cite{wolf1989read}:
\begin{equation}
    f_{\text{RSW}}(x,a)=\sin \left(\frac{\pi}{2x}\right)\left[1-a\log\left\{\sin\left(\frac{\pi}{2x}\right)\right\}\right]
    \label{eq:s-r-w}
\end{equation}
where $x=\frac{\theta-\theta_{\text{min}}}{\theta_{\text{max}}-\theta_{\text{min}}}$, $\theta$ and $a$ are the misorientation angle and shape parameter \cite{sarochawikasit2021grain} respectively, while $\theta_{\text{min}}$ and $\theta_{\text{max}}$ represent the periodicity range of the $\gamma_{[110]}(\theta)$ function (0 to $90^\circ$ in this case). 
These two parameters are obtained by fitting eq.\ \eqref{eq:s-r-w} to atomistic calculations.

In our two-dimensional continuum grain growth model, the misorientation is calculated as the angular offset between the projected crystal axes of the grains on both sides of the GB. The inclination is then taken as the angel between the GB plane normal, $\vec{n}$, and these axes.

\subsubsection{Grain boundary excess volume:}

The excess volume for symmetric tilt boundaries in W has been calculated using atomistic methods by Zheng et al.~\cite{zheng2023structure}. The values for $[110]$-tilt GB are given in Figure \ref{fig:GBexcess} normalized by the value of $a_0$ in W (3.15~\AA). In agreement with the Wolf model \cite{wolf1989correlation}, the dependence of $\delta V$ on $\theta$ closely mimics that of $\gamma_{\rm [110]}$ with $\theta$ (Fig.\ \ref{fig:GBEnergy}). As Fig.\ \ref{fig:GBexcess} shows, this dependence is non-monotonic and here we simply perform a linear interpolation between data points for a given value of $\theta$. The data also show that $\delta V$ is only a fraction of $a_0$, ranging between zero and 0.14.

Several studies have estimated the width of grain boundaries in metals to be in the 0.5-to-1.0-nm range \cite{prokoshkina2013grain,divinski2011recent}.
For that reason, the parameter $w$ in eq.\ \eqref{eq:width} is set to $2a_0$ here. This choice of $w$ justifies the approximation $\delta V\ll w$ adopted in Sec.\ \ref{sec:hjump}. 



\subsubsection{Grain boundary mobilities:}\label{sec:mob}

Grain boundary mobility is one of the most challenging aspects of polycrystal evolution. While it is reasonable to assume that bi-crystallography is the primary determinant of grain boundary mobility (through its five-dimensional relationship \cite{gottstein2001grain}), recent mobility calculations obtained by matching polycrystal evolution models to dynamic grain boundary evolution observations have yielded no definitive correlation between misorientation, inclination, and GB motion \cite{ZHANG2020211}. This is consistent with MD studies in Ni bicrystals \cite{OLMSTED20093704}, while in others, such as Al \cite{janssens2006computing,ZHOU20115997}, mobilities display a marked misorientation dependence. In the present study, the problem is compounded by the presence of deuterium, which may hinder grain boundary mobility relative to the pristine material condition.

To address these issues, here we use recent atomistic results of GB mobility in W \cite{mathew2022interstitial}. In particular, we use a definition of $\matr{M}$ based on the decomposition of a general grain boundary into a set of $m$ disconnections described by a structure vector $\vec{H}$ \cite{chen2019grain}:
\begin{equation}
    \matr{M} = \frac{f_0 A}{kT}\sum_{m}\left(\vec{H}^{m} \otimes \vec{H}^{m}\right) \exp\left(-\frac{Q^m_{a}}{kT}\right)
    \label{eq:mobtensor}
\end{equation}
where $f_0$ is a prefactor, $k$ is Boltzmann's constant, $\vec{H}=[b,h,0]$ ($b$ and $h$ are the magnitude of the disconnection's Burgers vector and step height respectively), $Q_a$ is the activation energy, and $A$ is again the GB area. In the current model, we take only the leading term of the sum in eq.\ \eqref{eq:mobtensor}, i.e., $m=1$. The parameters obtained by molecular dynamics simulations are listed in Table.~\ref{table:mobility_parameters}, where results for pure (pristine) W are given as well as for hydrogen-loaded GBs with H concentration to be 0.01 \AA$^{-2}$ (labeled with superindices `P' and `H', respectively).
\begin{table}[htb]
\centering
\parbox{\linewidth}{\centering 
\caption{Parameters used to define the mobility tensor in W for different misorientations along the [110] tilt axis. The superscript `P' denotes pristine W, while `H' refers to hydrogen-loaded GB \cite{mathew2022interstitial}.}
\label{table:mobility_parameters}
\vspace{5pt} 
}
\vspace{5pt}
\begin{tabular}{|c|c|c|c|c|c|c|c|}
\hline
$\theta$ [deg] & $f_0^{\text{P}}$ [s$^{-1}$] & $Q_a^{\text{P}}$ [eV nm$^{-1}$] & $f_0^{\text{H}}$ [s$^{-1}$] & $Q_a^{\text{H}}$ [eV nm$^{-1}$] & $h$ [\AA] & $b$ [\AA] & $A$ [nm$^2$] \\
\hline
26.52 & $4.0 \times 10^{17}$ & 1.99 & $5.6 \times 10^{14}$ & 1.68 & 1.58 & 0.8 & 29.20 \\
58.99 & $4.9 \times 10^{17}$ & 11.09 & $8.4 \times 10^{15}$ & 10.26 & 2.60 & 2.3 & 27.17 \\
65.47 & $1.0 \times 10^{16}$ & 6.87 & $3.3 \times 10^{12}$ & 6.31 & 2.54 & 2.0 & 30.92 \\
70.53 & $1.7 \times 10^{14}$ & 1.95 & $1.3 \times 10^{13}$ & 1.56 & 1.30 & 0.9 & 28.94 \\
77.88 & $2.0 \times 10^{15}$ & 2.58 & $1.0 \times 10^{14}$ & 2.55 & 2.66 & 1.5 & 42.62 \\
82.95 & $1.8 \times 10^{16}$ & 1.00 & $3.0 \times 10^{15}$ & 1.00 & 2.66 & 1.3 & 25.25 \\
86.62 & $2.3 \times 10^{18}$ & 1.11 & $3.1 \times 10^{16}$ & 1.18 & 2.66 & 1.0 & 29.32 \\
\hline
\end{tabular}
\end{table}

\begin{figure}[htbp]
    \centering
    \subfloat[Grain boundary mobility in pristine W\label{fig:M11P}]{
        \includegraphics[width=0.48\textwidth]{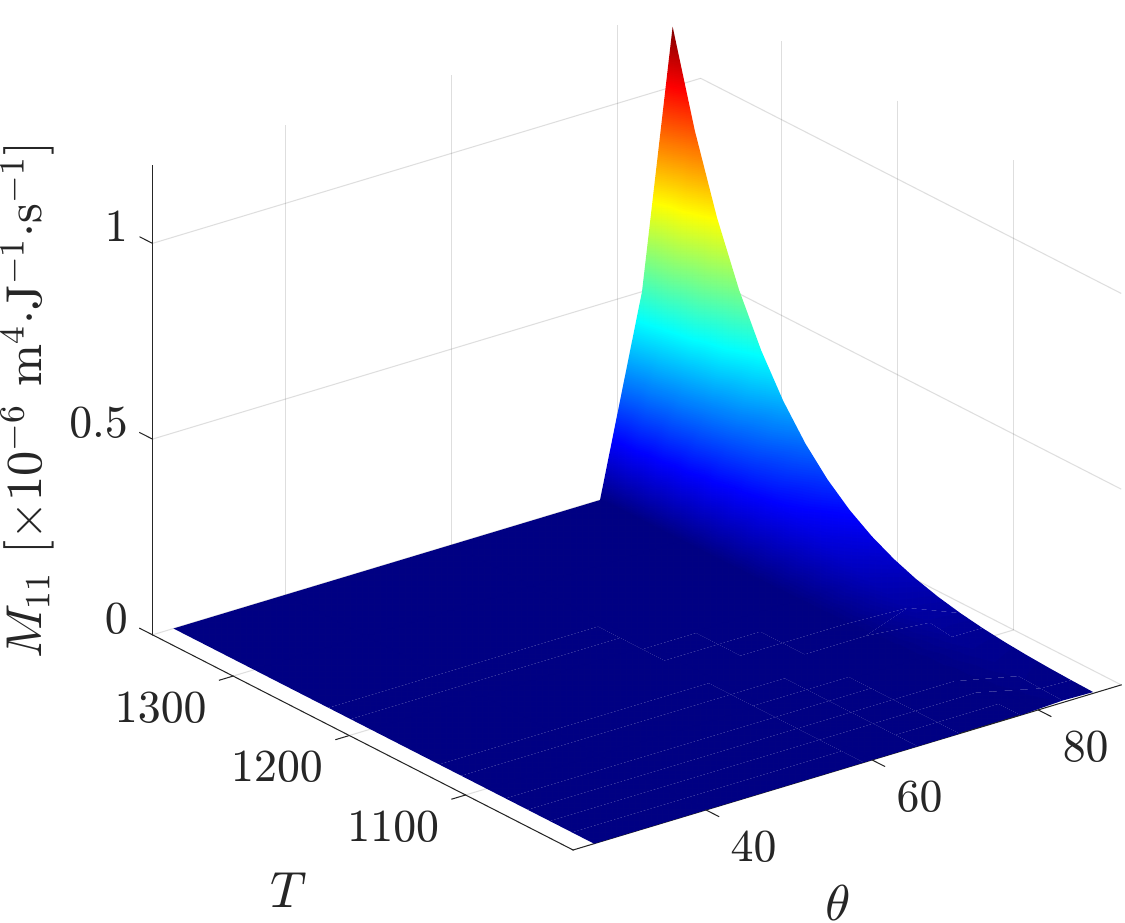}
    }
    \hfill
    \subfloat[Grain boundary mobility in H-loaded W\label{fig:M11H}]{
        \includegraphics[width=0.48\textwidth]{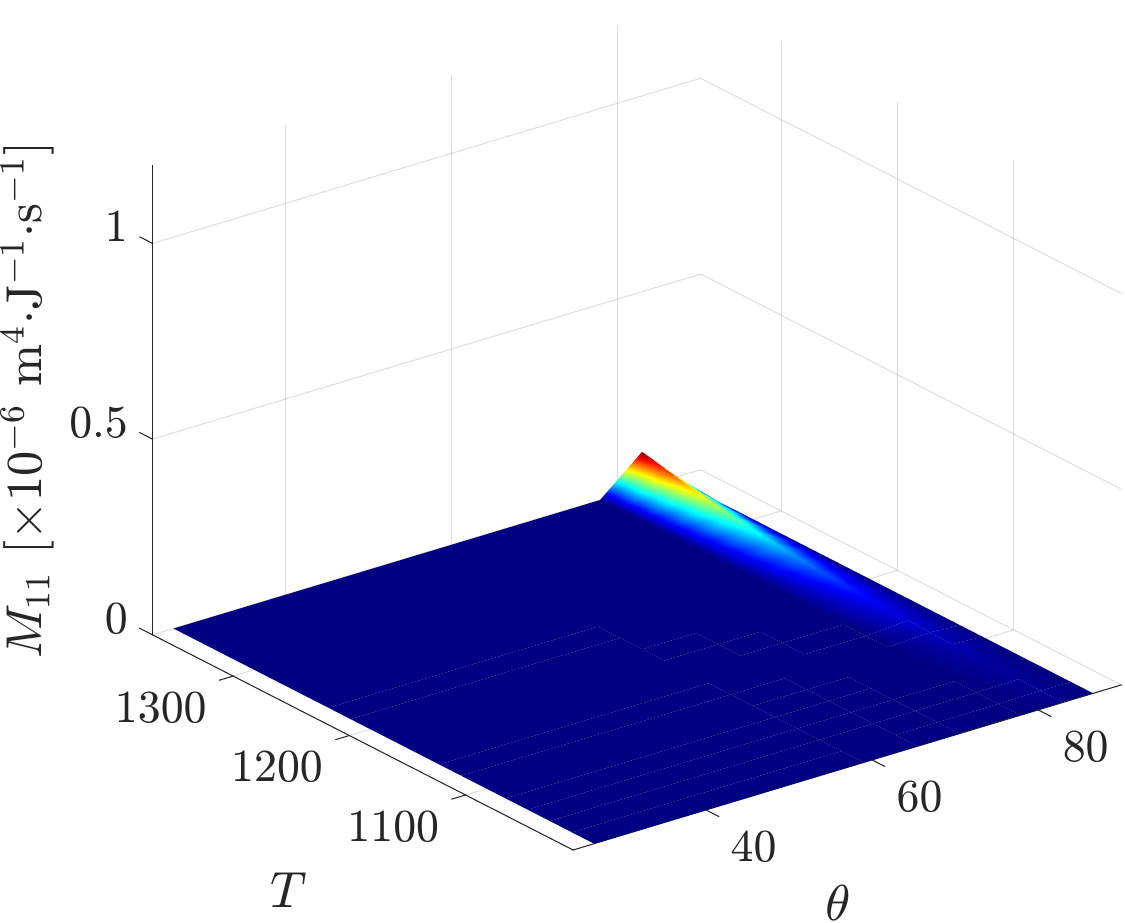}
    }
    \caption{The left and right hand side figures represent the GB mobility parameter $M_{11}$ without and with deuterium loaded respectively.}
    \label{fig:mobility_parameters_M11}
\end{figure}

Surface plots of $M_{11}$ for the pristine and hydrogen-loaded cases as a function of misorientation and temperature are given in Fig.~\ref{fig:mobility_parameters_M11}. As the figures show, GB mobility is virtually zero at temperatures below 900 K and for misorientations below 75$^\circ$. In the range where the mobilities are non-negligible, deuterium-loaded GB display values that are 10 times lower than those calculated for pristine W.

\subsubsection{Triple junction mobility:}\label{sec:tjmob}

Here we consider two regimes for TJ motion:
\begin{enumerate}
    \item The first regime is characterized by a ratio $\Lambda<1$, generically representing low TJ mobility. Accordingly, the kinetics is expected to be controlled by TJ --not GB-- motion, with  stresses arising from shear-coupled motion building up at unrelaxed TJs. As indicated above, beyond a certain stress threshold, such buildup may result in nucleation of twins or dislocations from the TJ into the grain. However, at present, this effect is not captured in the simulations.
    \item The second scenario involves $\Lambda>1$, representing conditions when GBs are expected to control the evolution of the microstructure. In this case, shear-coupled stresses are immediately relaxed by TJ motion, resulting in no emission of twins or dislocations.
\end{enumerate} 
In the results that will be presented below, we parametrically vary $\Lambda$ between 0.1 and 100, and study its effect on the resulting microstructures.  For clarity,  we use a value of $L$ equal to one in units such that $\Lambda$ reflects the non-dimensional ratio of mobilities in a direct way.

\section{Results}\label{res}

\subsection{Deuterium concentration profile and temperature distribution.}

\begin{figure}[htbp]
    \centering
    \subfloat[D profile \label{fig:concentrationProfile}]{
        \includegraphics[width=0.48\textwidth]{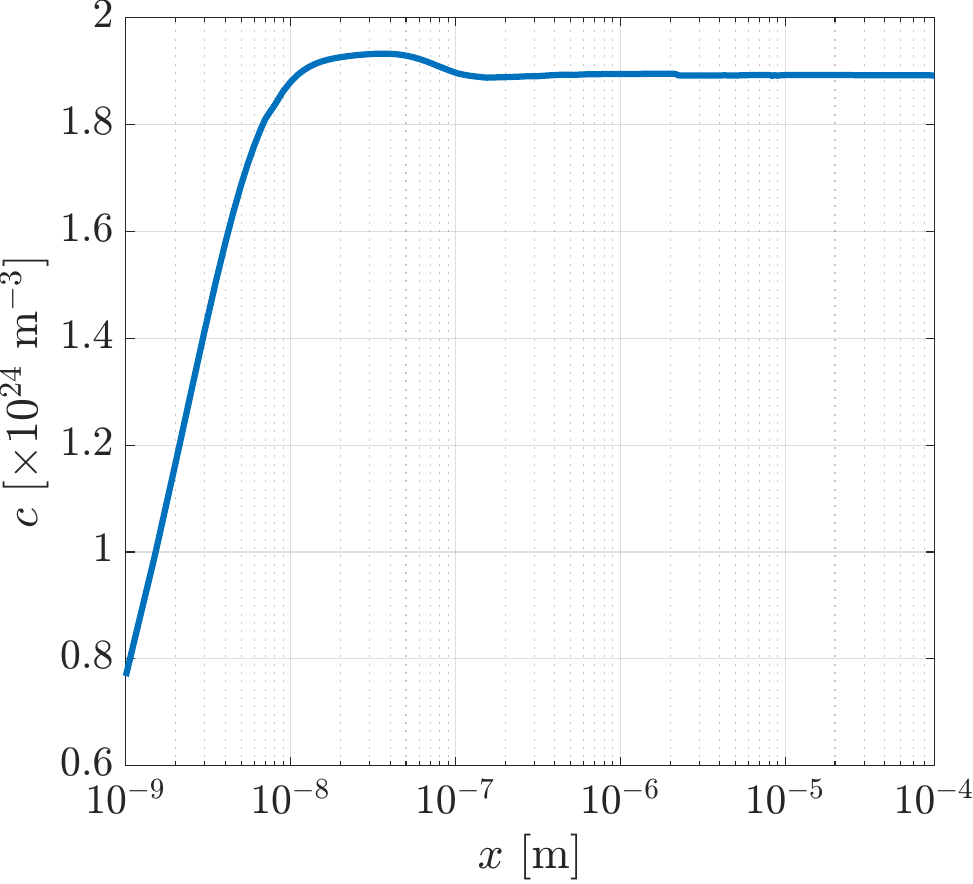}
    }
    \hfill
    \subfloat[Temperature profile \label{fig:temperatureProfile}]{
        \includegraphics[width=0.48\textwidth]{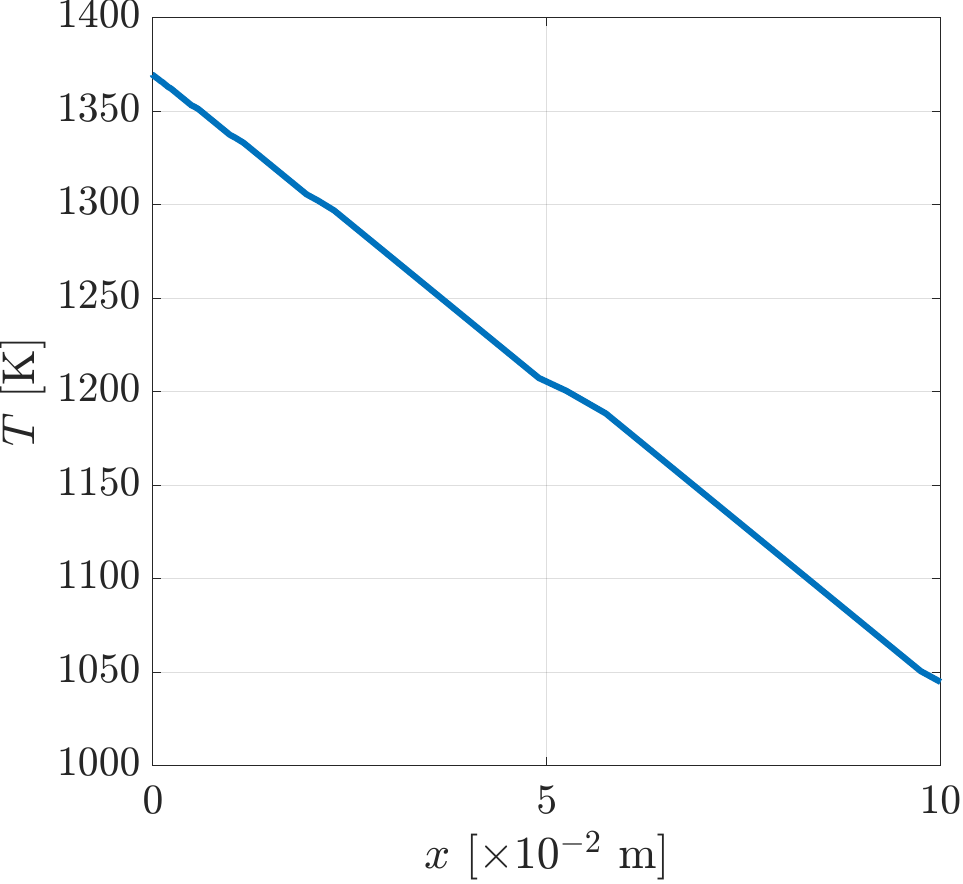}
    }
    \caption{{\protect\subref{fig:concentrationProfile}} Spatial concentration of deuterium, and {\protect\subref{fig:temperatureProfile}} temperature gradient as a function of depth of the W-based plasma facing component, after 10 cycles of 900 s of plasma exposure with a 60-s break, calculated by \texttt{Xolotl} for the location with peak plasma and heat fluxes, near the strike point. \label{xolotl}}
\end{figure}

Figure \ref{xolotl} shows depth profiles for the deuterium concentration and the temperature of the outer divertor target after exposure to 10 cycles of 900 s plasma with a 60-s break, as calculated by \texttt{Xolotl} for the location of peak plasma and heat fluxes at several locations near the strike point. 
Figure \ref{fig:concentrationProfile} reveals a flat steady-state deuterium concentration profile beyond the first 0.1 microns of material, at a level of $\approx1.9\times10^{24}$ m$^{-3}$. Thus, in our simulations we simply consider a uniform D distribution throughout the entire depth of the material (subject to the concentration jumps across GB discussed in Sec.~\ref{sec:hjump}). For its part, the temperature follows a decreasing linear profile from which a constant gradient of $3.2\times10^5$ K$\cdot$m$^{-1}$ can be extracted. In our simulations, we fix the temperature at the edge of the computational domain to 1370 K, and apply the constant gradient to set the temperature across it.

\subsection{Material microstructure and selection of the computational domain}\label{sec:micro}

\begin{figure}[htbp]
    \centering
\subfloat[\label{fig:FullEBSD}]{\includegraphics[width=0.6\textwidth]{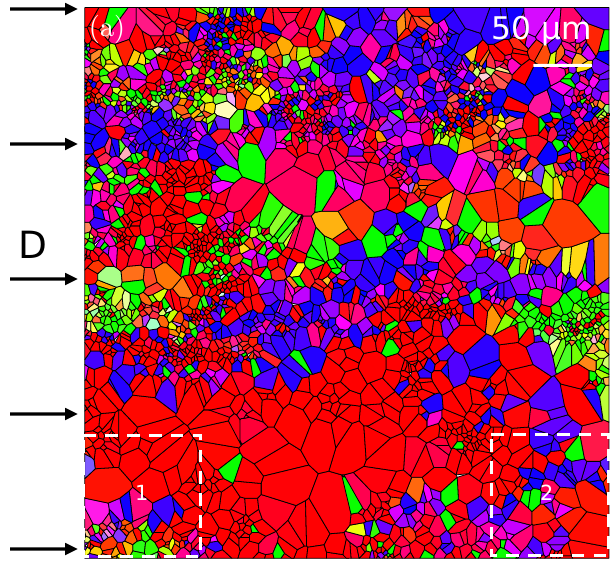}}
\subfloat[\label{fig:IPFLegend}]{
        \includegraphics[width=0.3\textwidth]{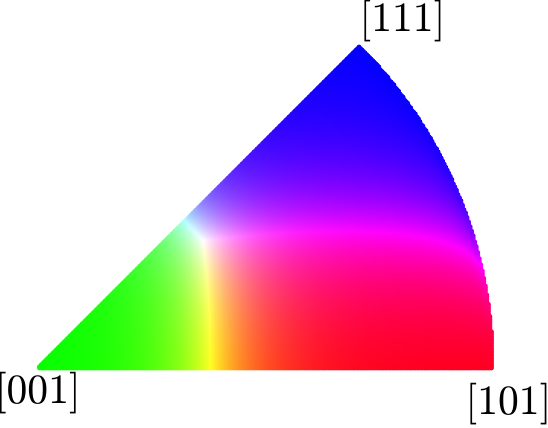}}\\
    \subfloat[Initial grain size distribution\label{fig:grainMisorientation1}]{
        \includegraphics[width=0.49\textwidth]{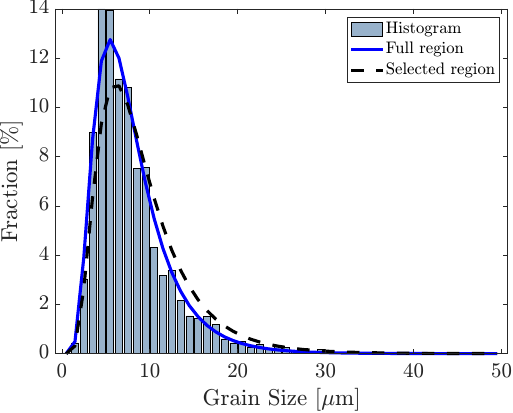}
    }
    \hfill
    \subfloat[Initial orientation distribution function\label{fig:grainSizeDistribution}]{
        \includegraphics[width=0.48\textwidth]{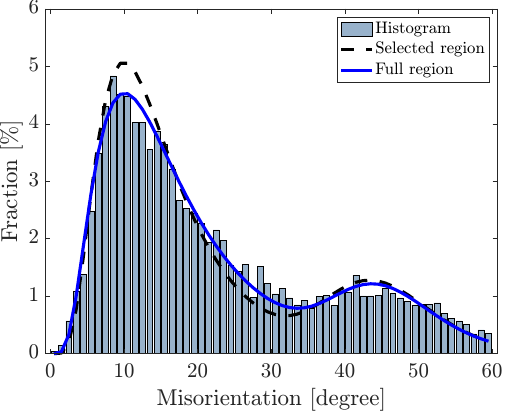}
    }
    \caption{
    {\protect\subref{fig:FullEBSD}} Full EBSD map of the as-fabricated W polycrystal projected along the 111 zone axis.
    {\protect\subref{fig:IPFLegend}} Color map of the crystal orientation in the unit triangle. The left-hand side is exposed to the deuterium plasma (marked by arrows).
    {\protect\subref{fig:grainMisorientation1}} Grain size histogram for full microstructure and computational subsets with associated log-normal fits.  
    {\protect\subref{fig:grainSizeDistribution}} Orientation distribution function for full microstructure and computational subsets with associated bimodal fits (inset '1' in Fig.\ \ref{fig:FullEBSD}).
    }
    \label{fig:grainSize}
\end{figure}

Next, we create synthetic two-dimensional microstructures from EBSD (electron backscattering diffraction) characterizations of pure W specimens obtained by the following sequence: (1) cold isostatic pressing at room temperature; (2) sintering at 1800$\sim$2200$^\circ$C; (3) hot rolling at 1400$\sim$1600$^\circ$C to an area reduction of approximately 80\%; (4) stress relief heat treatment at 900$^\circ$C for 20 min. Figure \ref{fig:FullEBSD} shows the `as-fabricated' microstructure colored according to the grain orientation as indicated in the attached standard triangle color map. The figure shows the surface exposed to D from the plasma (l.h.s.~of the image) and a scale marker for reference. Figure \ref{fig:IPFLegend} shows the color map in the stereographic triangle used to represent the crystal orientation in each grain. Based on the color map, the material appears to be textured with a slight predominance of crystal orientations near the [101] crystallographic axis (colored in red in the image).

The entire microstructure shown in Fig.\ \ref{fig:FullEBSD} contains 2994 grains and over 18,000 grain boundaries, which is presently beyond the computational capabilities of our vertex dynamics model. Instead, we use a subset of the configuration shown in the figure for the actual simulations. The actual microstructure employed in the simulations is highlighted by a white dashed box at the bottom left corner of Fig.\ \ref{fig:FullEBSD} (referred to as `inset 1'). This substructure contains approximately 115 grains and 584 grain boundaries, which makes it tractable computationally. However, to ensure that this subset is representative of the full set we conduct a statistical verification exercise comprising the grain size distribution and the misorientation distribution function (MDF), i.e., the normalized misorientation angle distribution. Figure \ref{fig:grainMisorientation1} shows a histogram with the initial grain size distribution of the microstructure shown in Fig.\ \ref{fig:FullEBSD}. The figure also shows a log-normal fit to the histogram data, from which an average grain size of 5.0$\sim$7.0 microns can be extracted. Importantly, the log-normal fit to the grain distribution extracted from the inset is also shown in the figure and found to be in reasonable agreement with that corresponding to the entire microstucture. 

The MDF is shown in Figure \ref{fig:grainSizeDistribution}. The MDF displays a bimodal shape, with a peak at misorientations of $\theta\approx10^\circ$ and another at $\approx45^\circ$. As in the case of the grain size distribution, the MDF is fitted to a dual log-normal composite function, corresponding to the entire and subset structures shown in Fig.\ \ref{fig:FullEBSD}, are in good agreement with one another. Combined, the results in Figs.\ \ref{fig:grainMisorientation1} and \ref{fig:grainSizeDistribution} provide evidence of the statistical equivalence between the full microstructure and the subsets chosen for computational efficiency. 

A similar analysis performed for the substructure highlighted by the white dashed box in the bottom right portion of the EBSD image (to be used for simulations described below in Sec.~\ref{sec:lowt}) led to the same conclusions.

\begin{figure}[htbp]
    \centering
    \subfloat[\label{fig:grainMisorientationIPF}]{
        \includegraphics[height=0.45\textwidth]{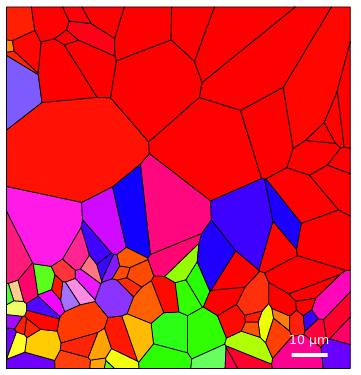}
    }
    \hfill
    \subfloat[\label{fig:grainMisorientation}]{
        \includegraphics[height=0.45\textwidth]{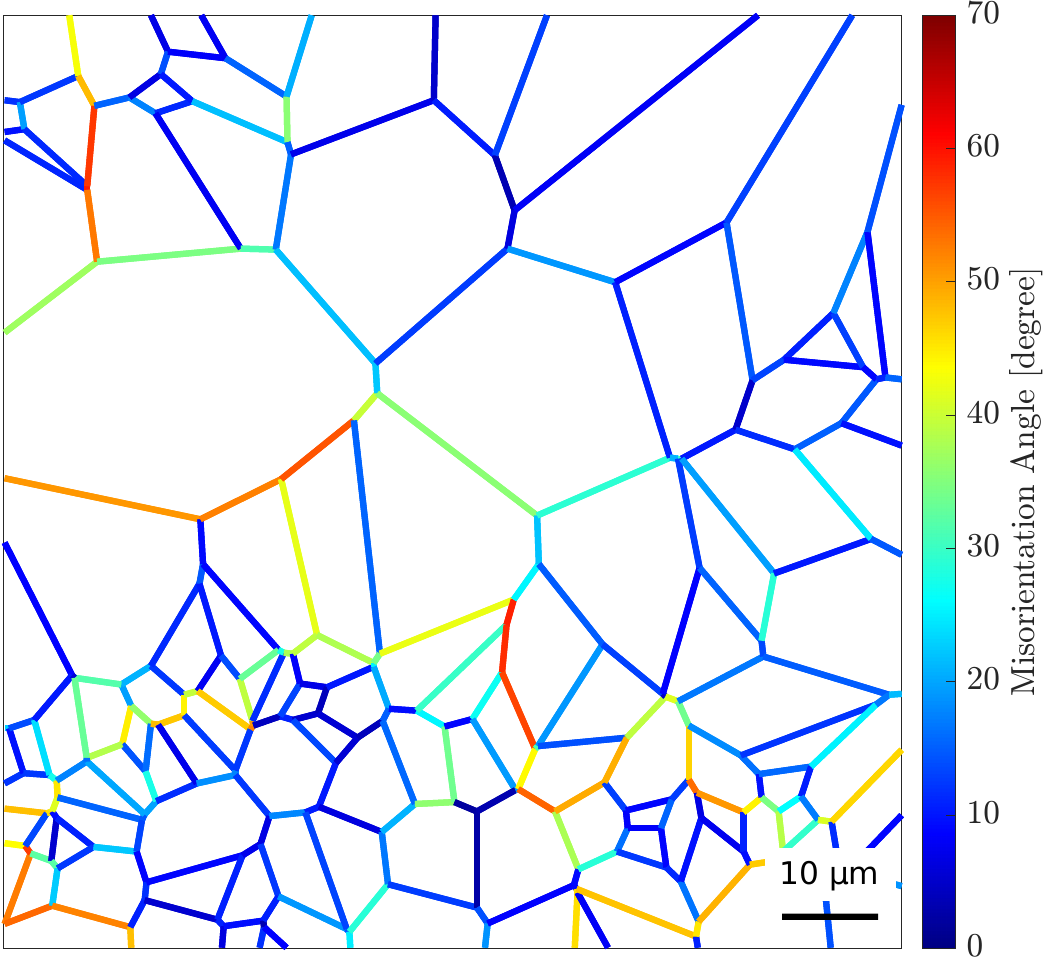}
    }
    \caption{
    {\protect\subref{fig:grainMisorientationIPF}} Grain crystal orientation according to the color map provided in Fig.\ \ref{fig:IPFLegend}.
    {\protect\subref{fig:grainMisorientation}} Grain boundary misorientation map.
    }
    \label{fig:Initial_Sample}
\end{figure}
Figure \ref{fig:grainMisorientationIPF} shows the grain substructure of subset `1' (also colored according to the map provided in Fig.\ \ref{fig:IPFLegend}), while Fig.\ \ref{fig:grainMisorientation} shows the corresponding grain boundary misorientation map. As the subfigure shows, the microstructure in subset `1' is dominated by low-angle grain boundaries ($<20^\circ$). Next, we carry out simulations of microstructure evolution in this subset with the fully-parameterized model.

\subsection{Grain growth simulations}

\subsubsection{Microstructure evolution:}

Here, we track the temporal evolution of the system and compute the fraction of transformed surface area by discretizing the entire grain domain into a $100\times100$ uniform grid. A given location within the grid is said to be transformed if a GB sweeps over it at any point in time. Each grid point is counted only once as transformed, regardless of whether it has been swept more than once. The fraction of the swept area, $f_A(t)$, is then calculated as the ratio of transformed grid points relative to the total number.

\begin{figure}[ht!]
\includegraphics[width=1.0\linewidth]{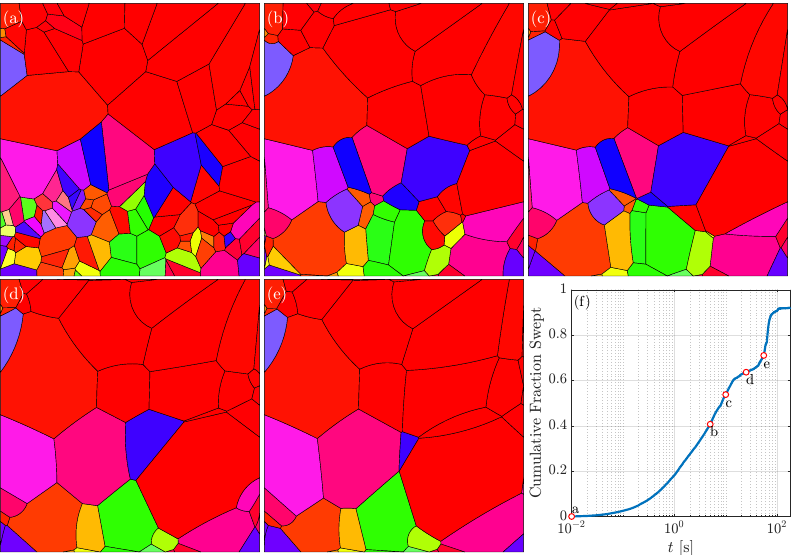}\hspace{-1.78em}%
\caption{\label{fig:grainGrowth}
(a)-(e) Different snapshots in time of the grain substructure in inset `1' (see Fig.\ \ref{fig:FullEBSD}). (f) Evolution of the transformed areal fraction with time from 0 to 100\%. The point in time of each of the snapshots in subfigures (a)-(e) is shown for reference.
}
\end{figure}

Figure \ref{fig:grainGrowth} shows a sequence of snapshots of the microstructure in subset `1' from $t=0$ until the maximum transformation is achieved. Visually, the smaller grains in the lower third sector of the image are seen to undergo rapid growth, relative to the rest of the substructure. As well, grains with crystal orientations close to the $[101]$ zone axis (colored red in the figure) appear to suffer a slower transformation. The evolution of $f_A(t)$ with time is shown in Figure \ref{fig:grainGrowth}f, showing saturation after 100 s of simulated time. The simulations predict that grain growth will begin instantaneously upon exposure of W to deuterium at the temperatures of the first wall (Fig.\ \ref{fig:grainGrowth}a). Rapid growth persists during the following 10 s of evolution (Figs.\ \ref{fig:grainGrowth}b-c), primarily by consumption of small grains by larger grains everywhere in the specimen. Subsequently, the system experiences slower growth (Figs.\ \ref{fig:grainGrowth}d-e), likely due to a `locked' microstructure limited by the slow growth of only one or very few specific grains. This manifests itself as a `bend' in the growth curve between 40 and 70 s in Fig.\ \ref{fig:grainGrowth}f. The system is seen to saturate at approximately 100 s of evolution, with 80\% of the total area swept.

\subsubsection{Quantitative analysis of the grain substructure}
The general observations from the system evolution presented in the prior section are that the microstructure coarsens by absorption of smaller grains by larger ones. Here we provide a more quantitative analysis of the main microstructural descriptors and their evolution during the grain growth process.

\begin{figure}[h!]
    \centering
    \subfloat[\label{fig:grainSizeHistogram}]{       \includegraphics[height=0.38\textwidth]{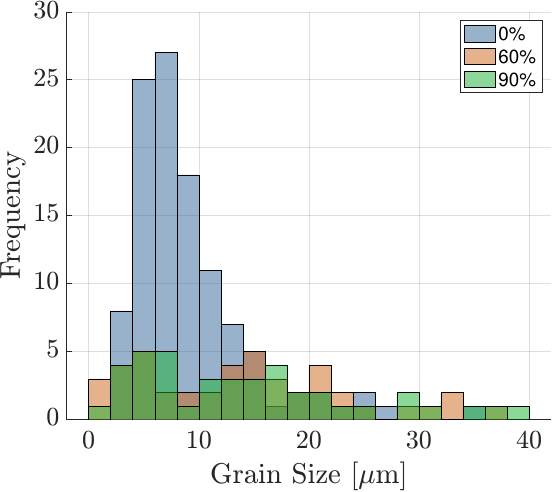}
    }
    \hfill
    \subfloat[\label{fig:grainSizeEvolution}]{      \includegraphics[height=0.38\textwidth]{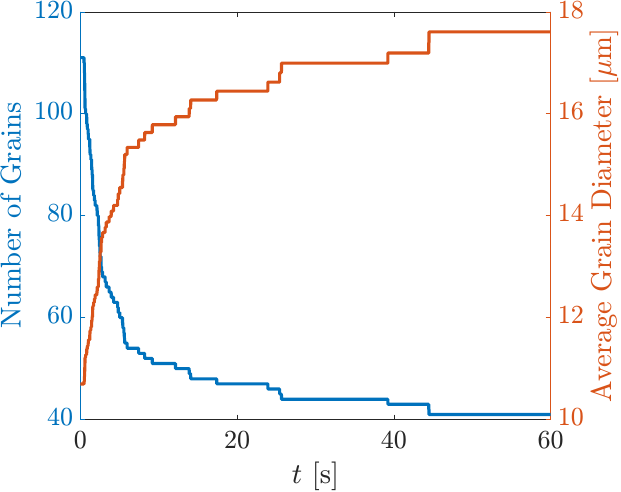}
    }
    \caption{
    {\protect\subref{fig:grainSizeHistogram}} Grain size distribution at $0$, $60$, and $90\%$ completion of the grain growth process.
    {\protect\subref{fig:grainSizeEvolution}} Number of grains and average grain diameter as a function time.
    }
    \label{fig:grainSizeAnalysis}
\end{figure}

Figure \ref{fig:grainSizeAnalysis} contains information about the evolution of the grain size with time. The grain size distribution at three stages during the growth process, 0, 60, and 90\% completed, is shown in Figure \ref{fig:grainSizeHistogram}. Clearly, the distributions become more uniform, deviating from the original log-normal. This is accompanied by coarsening, as shown in Fig.\ \ref{fig:grainSizeEvolution}, where a monotonic reduction in the total number of grains from the original 115 to merely 41 of them. Likewise, the average grain size increases from 10.5 to $\approx18$ microns.

\begin{figure}[H]
\centering
\includegraphics[width=0.5\linewidth]{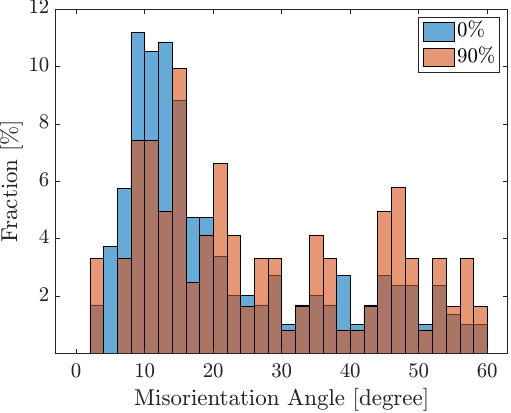}
\caption{\label{fig:misorientationAnalysis}
Misorientation distributions before and after grain growth. 
}
\end{figure}

\begin{figure}[H]
    \centering
    \subfloat[\label{fig:percentageMisorienGG_grouped}]{        \includegraphics[width=0.48\textwidth]{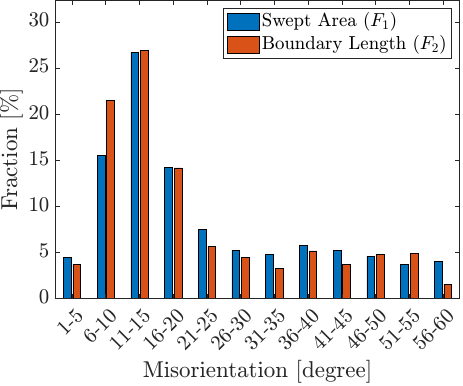}
    }
    \hfill
    \subfloat[\label{fig:percentageMisorienGG_ratio}]{\includegraphics[width=0.49\textwidth]{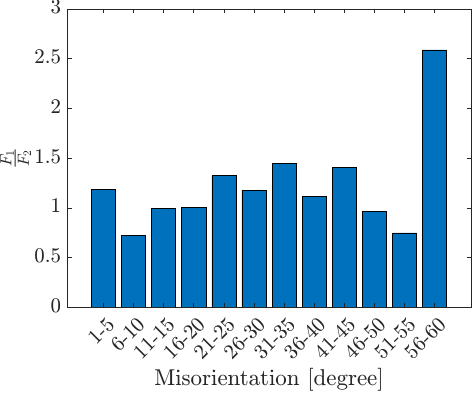}
    }
    \caption{
    {\protect\subref{fig:percentageMisorienGG_grouped}} Fraction of swept area and initial total GB length for each GB misorientation bin.
    {\protect\subref{fig:percentageMisorienGG_ratio}} Relative importance of each GB misorientation in the grain growth process.
    }
    \label{fig:percentageMisorienGG}
\end{figure}

The change in the misorientation distribution function between $f_A=0$ and $f_A=0.9$ is given in Figure \ref{fig:misorientationAnalysis}. As the figure shows, while the changes are not drastic, a depletion in the relative proportion of LAGB in favor of higher angle ones can be clearly observed. A priori, this may suggest that highly mobile GB are activated towards eliminating grains with a high proportion of LAGB (which are less mobile according to Fig.\ \ref{fig:mobility_parameters_M11}). To elucidate this point, however, a more appropriate analysis is to directly quantify the relative participation of each GB in the entire grain growth process. To that effect, in Figure \ref{fig:percentageMisorienGG_grouped} we show the distribution of misorientations at the beginning of the simulations ($F_2$, equivalent to Fig.\ \ref{fig:grainMisorientation1}) next to their contribution to the swept area fraction, $F_1$. 
The relative weight of each misorientation bin in the total grain growth process is then established by obtaining the ratio between $F_1$ and $F_2$, which is shown in Figure \ref{fig:percentageMisorienGG_ratio}. It is clear that HAGB, $>50^\circ$, contribute the most proportionally to grain growth, almost a factor of two more than any other GB misorientation. This points to GB mobility as the controlling factor.

\subsubsection{Low temperature grain growth simulations:}\label{sec:lowt}
\begin{figure}[H]
\centering
\includegraphics[width=0.5\linewidth]{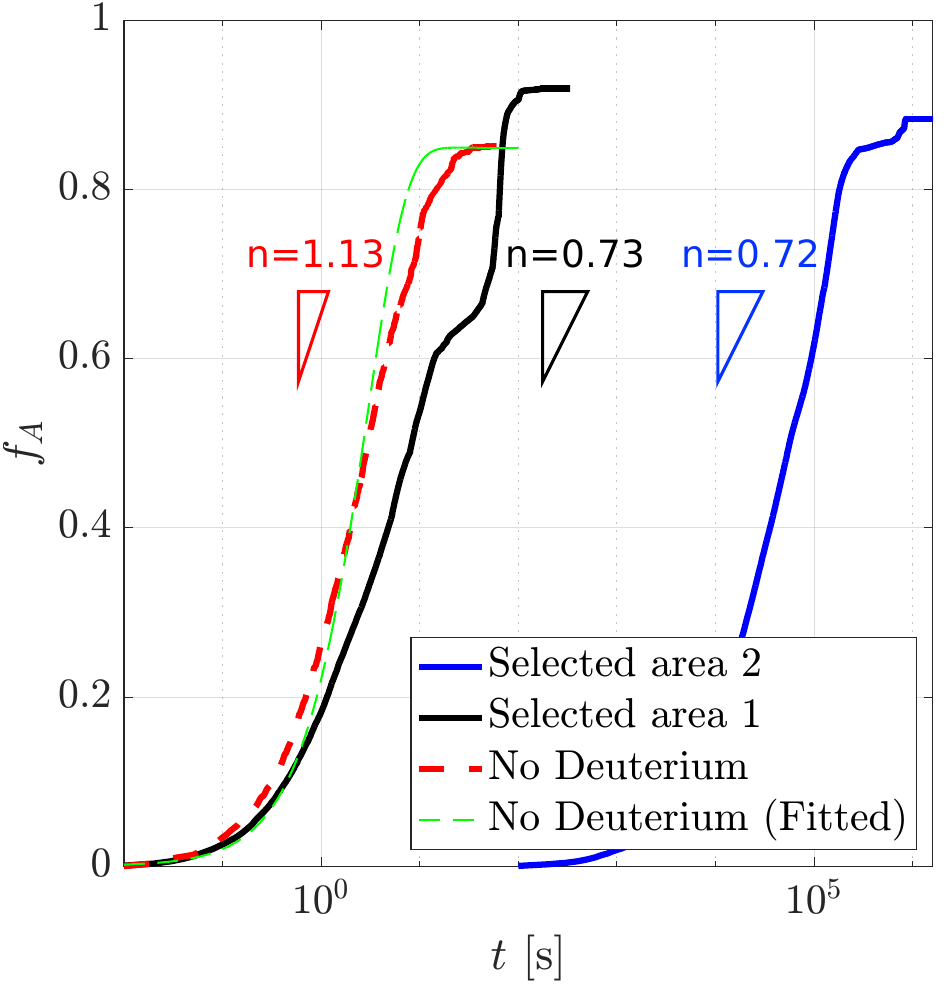}
\caption{\label{fig:evolution1}
Evolution of the swept areal fraction as a function of time for two cases at 1400 K (with and without D), and one case (with D) at 1000 K. The cases with D correspond to the regions in Fig.\ \ref{fig:FullEBSD} labeled as `1' and `2', respectively. 
The Avrami exponents in each case have been calculated via fits to eq.\ \eqref{eq:jmak}.
The dashed green line represents the fitted line for without D case at 1400K. 
}
\end{figure}

Last, we carry out simulations of grain growth at 1000 K for the microstructure contained in inset `2' shown in Fig.\ \ref{fig:FullEBSD}. Such microstructure satisfies the statistical equivalence conditions of grain size and grain boundary misorientation distributions as described in Sec.\ \ref{sec:micro}, but we use it instead of inset `1' for statistical variation. The results are shown in Figure \ref{fig:evolution1}, where the fraction of swept area is shown as a function of time. The microstructure begins to evolve after 3 s of clock time, but its transformation is not completed until after $10^6$ s (approximately 11.5 days). The curve for the microstructure in inset `1' (adjacent to the plasma) is also shown for comparison, together with the case for pristine W where only capillary forces (GB curvature) are at play. Clearly, temperature is the most significant factor for microstructural evolution and grain boundary activation.

\section{Discussion} \label{disc}

\subsection{Model assumptions and simplifications}

\begin{itemize}
\item \emph{Triple junction mobility}: Here, we have assumed that TJ mobility is much higher than GB mobility such that triple junctions immediately adopt their equilibrium configuration in response to GB motion. TJ mobility is a challenging parameter to obtain experimentally, as triple-junctions are of course 2D representations of 3D structures \cite{gottstein2005triple}. There are very few modeling works \cite{upmanyu1999triple,upmanyu2002molecular}.  Recent works have established the effect of TJ mobility on grain growth (described by a growth scaling law of $\sim$$t^{n}$) when considered in conjunction with other parameters \cite{streitenberger2014triple,zollner2016grain}. In particular, these are expected values of $n$ depnding on what controls grain growth kinetics:
\begin{itemize}
    \item [(i)] Controlled only by GB mobility: a time exponent of $n=1/2$ is to be expected.
    \item [(ii)] Controlled by the energy of the grain boundaries together with the mobility of the triple junctions: $n=1$.
     \item [(iii)] Controlled by the energy of the triple junctions together with the mobility of the grain boundaries: $n=1/3$.
     \item[(iv)] Finally, when the kinetics is solely controlled by the triple junction mobility, $n=1/2$ again.
\end{itemize}
Of course, other factors such as defects, dislocations, impurities, etc, lead to deviations from these expected values. In our study, our model runs under conditions (i) and (ii) above, which is consistent with the Avrami exponents that will be discussed in Sec.\ \ref{}.   

    \item \emph{Two-dimensional model}: The present model considers polycrystals as 2D systems representing an axis zone projection of full 3D microstructures. The model is suitable for reconstructions of polycrystals from EBSD maps provided that the Euler angles for all grains and the zone axis are known (both of which are the case here). The implications of representing 3D polycrystals in 2D have been discussed in our previous work \cite{mcelfresh2023initial}. Suffice it to say that Potts Monte Carlo simulations in 2D and 3D have shown that $n=1/2$ (representative of 2D kinetics) can be considered to be the asymptotic long-time kinetic growth exponent even for three-dimensional grain growth \cite{anderson1989computer}.

    \item \emph{Selection of $[110]$ tilt axis}: The reason to consider GB properties for $[110]$ tilt boundaries in this study is simply that that orientation is the only one for which GB energies and mobilities are jointly available. It is not clear at this point what effect, if any, this has on grain growth and microstructural evolution. Given that GB energies for $[110]$ tilt boundaries are quantitatively similar to other crystal axes (see, e.g., ref.\ \cite{frolov2018structures}), and that the dependence of GB mobilities on temperature and misorientation are not expected to differ substantially from those used here (see ref.\ \cite{mathew2022interstitial}), we speculate that the choice made here is reaosnably representative of all GB tilt orientations. It has been shown, however, that the structure of any GB with the misorientation axes $[110]$ (and $[111]$) may be represented by a limited number of basic structural units \cite{urazaliev2021structure}, such that $[110]$ tilt boundaries are an ideal choice for general grain growth simulations.
    
    \item \emph{Influence of plasticity and residual stresses}: As indicated in Sec.\ \ref{sec:intro}, fusion-grade tungsten typically involves warm or hot rolling to bring the DBTT to engineering acceptable levels. This will inevitably lead to the buildup of dislocation density and the onset of residual stresses. In our previous work \cite{mcelfresh2023initial}, we have shown how to capture the effect of plasticity in driving grain growth processes, and, per eq.\ \eqref{pepei}, the model is intrinsically capable of incorporating stresses. However, the microstructure maps featured in Fig.\ \ref{fig:FullEBSD} do not include any information beyond grain orientations and sizes. Thus, the present work does not intend to be a numerical analysis of the response of realistic tungsten to fusion conditions, but a study focused on the effects of deuterium exposure to pure W polycrystals.

\end{itemize}

Despite the simplification and assumptions above, the model has value in assessing the susceptibility of monolithic W to grain growth under ARC operational conditions, as well as estimating the time scales over which it takes place, as will be shown below.

\subsection{Avrami exponents and dimensionality of the kinetics}\label{}

To assess the kinetics of grain growth in an analytical fashion, next we fit the curves in Fig.\ \ref{fig:misorientationAnalysis} to a modified Johnson-Mehl-Avrami-Kolmogorov (JMAK) model \cite{fanfoni1998johnson,shirzad2023critical}, i.e., where the fraction of transformed area can be represented by the following expression:
\begin{equation}
    f_A = 1 - \exp\left(-a(t - t_0)^n\right)
    \label{eq:jmak}
\end{equation}
where $a$ is a fitting constant, $n$ is the the Avrami exponent (which describes the dimensionality of the governing mechanisms behind grain growth), and $t_0$ represents the incubation time before the start of the transformation. For a grain growth process purely defined by GB motion, a value of $n\approx2$ is expected. As indicated in Fig.\ \ref{fig:misorientationAnalysis}, the Avrami exponents extracted from our simulations are 0.73 for the cases containing deuterium and 1.13 for the case without. 
These values are in excellent agreement with experimentally-measured values.
Alfonso et al.~\cite{alfonso2014recrystallization} found a value of $n=1.1$ in their study of recrystallization kinetics of warm-rolled tungsten, while Wang et al.~\cite{wang2017effects} found values between 0.78 and 1.0 in heavily pre-deformed W at intermediate temperatures. Finally, Shah et al.~\cite{shah2021recrystallization} found an average value of 1.4 in their study of H-plasma exposed W specimens. Again, this is remarkable agreement with our simulation results given the differences and assumptions pertaining to the model.

Furthermore, tests with varying temperatures allow for the determination of the activation energy for grain growth, which, for H-loaded samples, was estimated to be approximately 4.5 eV \cite{shah2021recrystallization}. Referring to this value to the activation energies GB listed in Table \ref{table:mobility_parameters} suggests that the evolution of the system is controlled by GB with misorientation angles of $\approx78^\circ$. 
This is consistent with the results presented in Fig.\ \ref{fig:percentageMisorienGG_ratio}, where HAGB clearly contribute the most to the entire grain growth process.

\subsection{Effect of deuterium on grain growth}

As implemented in the model, D has two main effects on microstructural evolution. The first one is preferential segregation towards grain boundaries, which results in concentration jumps that scale with misorientation according to eq.\ \eqref{eq:width} and Fig.\ \ref{fig:GBexcess}. This leads to differential D accumulation across GB, which, in turn, results in driving forces for GB evolution.
The second is a reduction in GB mobility due to an increased activation energy as indicated in Table \ref{table:mobility_parameters}. 

From the simulations, it is clear that D has a delaying effect on GB kinetics, as manifested by the lower Avrami exponent of 0.72 compared to a value >1.0 for pristine W. Interestingly, there is no effect on the incubation time (i.e., the time at which the transformation commences), which is apparently controlled only by temperature.

\subsection{Implications for plasma-facing armor}

Ultimately, our simulations suggest that monolithic tungsten fabricated using conventional techniques will be highly susceptible to grain growth in the presence of any driving force at temperatures above 1000$^\circ$C. Here, we have considered GB curvature (probably established during grain deformation processes due to hot rolling) and differential deuterium accumulation at GB (assumed to occur due to excess atomic volume in the vicinity of boundaries) as the main driving forces in our study. We have shown that those alone are sufficient to drive microstructural evolution at 1400 K to >80\% transformed fraction within less than 100 s. Decreasing the temperature to 1000 K, delays the onset of grain growth by four orders of magnitude in time, clearly highlighting the critical role of temperature in driving internal changes.

However, one must also consider that, in general, pre-strained tungsten will contain a relatively high density of residual dislocations. Since plastic deformation is highly-dependent on crystal orientation, it is expected that large differential dislocation density accumulation around GB will result in large driving forces for microstructural evolution, leading to even faster transformations. A possible technique to delay grain growth and stabilize pre-deformed W microstrutures is to take advantage of Zener pinning by alloying with selected solutes or seeding with second-phase particles. In fusion-grade W, promising systems include W-Re \cite{he2006microstructural,du2024comparative,shi2024effects,tsuchida2018recrystallization}, W-K \cite{tan2024dbtt}, W-Re-K \cite{gietl2022neutron,nogami2020tungsten}, or W infiltrated with lanthanide oxide particles \cite{mabuchi1997deformation,munoz2011la2o3}. In most of these cases, it is indeed found that W alloys display a higher recrystallization temperature relative to pure W.

Finally, under irradiation, two competing effects can be envisaged to appear. First, an extra driving force is introduced in the form of irradiation defects accumulating in different proportions on both sides of the GB. This could presumably accelerate microstructural evolution in the same manner as GB curvature, plastic deformation, and deuterium exposure do. At the same time, however, irradiation defects are known to strongly interact with grain boundaries, leading to complex structures \cite{he2024complex} which may pin GB and hinder the onset of grain growth or alter the GB mobility\cite{borovikov2012influence}. From the limited experimental evidence currently available \cite{fukuda2012effects,gietl2022neutron,luo2023effect}, it appears that irradiation damage also leads to delayed microstructural evolution processes in comparison with unirradiated W. However, we leave the effects of irradiation on grain growth for future studies.

\section{Conclusions} \label{conc}

In summary, we have developed a vertex dynamics model in 2D to assess the grain growth kinetics of deuterium-exposed polycrystalline tungsten. The model tracks the motion of grain boundaries under the effect of driving forces stemming from grain boundary curvature and deuterium concentration jumps. All the model parameters have been obtained from atomistic simulations, including grain boundary energies, mobilities, excess volumes, and deuterium solution energies. The simulations assume saturation conditions for D-exposed tungsten as obtained from integrated plasma (\texttt{UEDGE} code) -material (\texttt{F-TRIDYN} and \texttt{Xolotl}) modeling of the outer divertor target exposed to 10 cycles of 900 s of plasma exposure with 60-s breaks, at locations of peak plasma and heat fluxes near the strike point.

Our main findings are listed below:
\begin{itemize}
    \item The initial material microstructure has been analyzed statistically from two-dimensional EBSD maps. We find that the grain size distribution follows a log-normal distribution with an average grain size of approximately 11 microns. The grain boundary misorientation function is bimodal, with one peak at 10$^\circ$ and another at 45$^\circ$.
    \item In the plasma-facing region (first 100 microns, 1400 K), grain growth commences immediately upon deuterium saturation of the microstructure, taking less than 100 s to complete. The average grain size is seen to grow from 11 to 18 microns.
    \item We find that high-angle grain boundaries ($\approx60^\circ$) contribute proportionally the most to grain growth. 
    \item Avrami exponents of microstructural evolution laws range between 0.72 for the case of D-exposed tungsten and 1.2 for unexposed W. These are in very good agreement with experimental measurements.
    \item Temperature has a drastic impact on grain growth kinetics, delaying the incubation time from $1.0$ to $10^4$ s when the temperature is reduced from 1400 to 1000 K.
    \item Deuterium has a pinning effect on grain boundaries, reducing their mobility and delaying the onset and progression of grain growth. 
\end{itemize}

\section*{Acknowledgments}

This work was funded under the INFUSE program, a DOE-FES public-private partnership, under contract no.~DE-SC0024662 between the University of California Los Angeles, the University of Tennessee-Knoxville, and Commonwealth Fusion Systems.
The authors also acknowledge support from DOE-FES under contracts DE-SC0018410 (J.Y.~and J.M.) and DE-SC0023180 (S.B., A.L., and B.D.W.). N.M. gratefully acknowledge support from the U.S. DOE, Office of Science, Office of Fusion Energy Sciences, and Office of Advanced Scientific Computing Research through the Scientific Discovery through Advanced Computing (SciDAC) project on Plasma-Surface Interactions (Award No. DESC0008875).
We thank Shuhei Nogami from ALMT for providing the tungsten material and Blake Emad at University of North Texas for performing the EBSD analysis on the tungsten specimens. This manuscript has been assigned LA-UR-25-27324.

\section*{References}
\bibliographystyle{iopart-num}
\bibliography{refs-jaime,refs-from-RXpaper,mybib,cas-refs,Biblio}

\end{document}